\documentclass[conference]{IEEEtran}
\IEEEoverridecommandlockouts
\usepackage{cite}
\usepackage{amsmath,amssymb,amsfonts}
\usepackage{algorithmic}
\usepackage{graphicx}
\usepackage{textcomp}
\usepackage{xcolor}
\def\BibTeX{{\rm B\kern-.05em{\sc i\kern-.025em b}\kern-.08em
    T\kern-.1667em\lower.7ex\hbox{E}\kern-.125emX}}

\newcommand{\mycomment}[1]{}

\begin{document}

\title{Consider an Applications-First Approach for PDC\\
}

\author{\IEEEauthorblockN{Michelle Mills Strout}
\IEEEauthorblockA{
\textit{Hewlett Packard Enterprise and University of Arizona}\\
michelle.strout@hpe.com}
}

\maketitle

\begin{abstract}
    
I propose an applications-first approach for adjusting how 
parallel and distributed computing concepts are incorporated into curricula. 
By focusing
on practical applications that leverage parallelism and distributed
systems, this approach aims to make these complex topics more accessible
and engaging for both CS and non-CS majors.
An applications-first approach demonstrates the advantages of parallel and 
distributed computing in solving real-world problems while building practical 
experience and skills before delving into theoretical concepts. This could 
potentially broaden the appeal and retention of these concepts.
I highlight some example application-centric efforts, and 
conclude with questions that could be investigated in the service of 
exploring applications-first approaches.

\end{abstract}


\section{Introduction}

The field of parallel and distributed computing (PDC) has become
increasingly crucial in today's computational landscape. However,
traditional approaches to teaching PDC often present significant
barriers to entry, particularly for non-CS majors.
For example, the ACM/IEEE curriculum for PDC recommends numerous
pre-requisites for the PDC knowledge area~\cite{acmieee}.
In this abstract for an EduHPC lightening talk, I propose
an applications-first approach to teaching PDC concepts, aiming to make
the field more accessible, appealing to, and therefore more effective
for a broader range of CS 
and non-CS students.

A lot of research has been done to determine how to leverage parallel 
hardware most effectively. This focus in the research has shaped
the focus of PDC curricula.  The emphasis often lies in understanding the 
intricate details required to achieve performance and writing PDC code from scratch.
However, this somewhat ignores the fact that many useful
PDC applications already exist and are used in production.  Additionally
many serial Python codes exist that CS and non-CS students
could do more with if such apps
were made parallel and distributed.  An applications-first
approach to PDC curricula would focus more on effectively executing
and extending these existing codes for HPC environments versus
squeezing every little bit of performance out of an application written
from scratch.

Currently, upper-division and graduate-level PDC courses typically
require substantial prerequisites, including advanced programming
skills, understanding of parallel architecture, and completion of data
structures and algorithms courses. This requirement structure often
limits access to non-CS majors and may deter some CS students as well.
While there have been attempts to create PDC courses catering to non-CS
majors, these often still maintain heavy prerequisite requirements and,
in some cases, have been discontinued 
(e.g. the USC Viterbi program~\cite{uscviterbi}).

Even for CS majors, PDC courses can be daunting.
As an undergraduate in the 1990s, I took a parallel programming class at 
UC, San Diego.  It was a fun class and introduced me to the world of High 
Performance and Scientific Computing, which I continue to enjoy and have 
built my career upon.  However, the Matrix-Matrix multiply assignment in 
MPI scarred me for life because even though I was the only one in the class 
that got it to execute in parallel and produce an answer, it had 
significant slowdown.  

What happened with that MM assignment is that it took a lot of time and 
code just to parallelize over rows and get the communication working.  The 
focus was on being able to implement parallelism from scratch versus
being able to identify and understand where parallelism is effective strategy.
In 2008 while teaching a CS3 data structures
course with 150 students, I received analogous feedback from students indicating they spent
most of their time on programming assignments writing the code to parse the
input versus learning the key concepts we were discussing in lecture.
In the PDC setting, current courses appear to be focusing more student time
on developing new applications and managing performance details than
understanding why HPC is so important and key parallelism concepts.

One of the primary concepts in PDC is that it enables computation
to be the fourth pillar of science, engineering, and even the
humanities through data analytics and simulation.
Some people like yours truly enjoy the complexity and want to dive
into the challenge of understanding all of the details of PDC.
So much so they are willing to wade through all that complexity before
they get to the interesting applications.
However, not enough people appear to enjoy the common approach
of teaching PDC in an underlying principles first way that involves
reimplementing smaller kernels such as MM
and other linear algebra algorithms as is indicated by
the need for programs to increase the size of the HPC workforce~\cite{HPCworkforce}.

\section{The Applications-First Approach}

An applications-first approach seeks to address these challenges by
introducing students to PDC concepts through the lens of practical
applications. Such a method would enable students to first understand the
advantages of parallelism and distributed computing by seeing their
impact on applications they find interesting or relevant to their fields
of study. Once students grasp the benefits, the underlying parallel
and distributed concepts could
be introduced more effectively.

Students should have the opportunity
to run existing parallel and distributed applications to (1) see the 
performance and scaling advantages they provide and
(2) observe those advantages being applied to real world problems.
These applications could be used to analyze
data, perform simulations, or solve problems that students find
interesting or relevant to their majors, whether in CS or other fields.

For non-CS majors, the focus would be on understanding how to use and
modify existing parallel and distributed applications, rather than
building them from scratch. This approach allows students to engage with
PDC concepts without requiring extensive programming prerequisites. CS
majors, on the other hand, could benefit from seeing practical
implementations before diving into the theoretical underpinnings,
potentially increasing their engagement with the field.

\section{Examples and Implementation}

Several communities and projects are already embracing aspects of the
applications-first approach.

The student cluster competitions are an effective implementation of this 
approach that focus on the applications and then tuning the performance
of those applications.  For CS students this is ideal.
However, for non-CS students, a more important next step after getting
applications running effectively in a parallel environment would then
be to add functionality to the applications.
Application-first PDC courses could cover how to use
existing applications, make modifications when performance issues arise,
extend existing applications, teach enough about parallel principles to categorize
different ways applications leverage those principles,
and eventually develop new parallel and distributed applications.

The Chapel parallel programming language~\cite{chapel-lang} exemplifies 
the applications-first approach. Chapel enables applications developed on 
laptops to be minimally adjusted for execution on parallel and distributed computing hardware, 
significantly reducing the time for collaborators to add functionality. 
For instance, Prof \'Eric Laurendau reported a reduction from a couple of years to about 
3 months in 
development time when masters students added new physics to the 
CHAMPS Computational Fluid Dynamics 
framework~\cite{laurendeau2021hpc, parenteau2021development}.
Additionally, the Chapel team conducts monthly meetings with educators to
 incorporate real-world Chapel applications and examples into teaching 
 materials\footnote{https://github.com/chapel-lang/ChapelExamplesAndTeachingMaterials}. 
 The Chapel team further eases the learning curve through various initiatives: 
 Python-to-Chapel transition blog posts for existing applications~\cite{corrodo2024navier}, 
 GitHub CodeSpaces with examples\footnote{https://chapel-lang.org/tryit-codespaces.html}, 
 installation assistance for various environments, 
 and ChapelCon\footnote{https://chapel-lang.org/ChapelCon.html},
 a conference showcasing user applications.


There are various other approaches that help ease the parallel learning curve.
For example, the development of X+CS curriculum~\cite{katz2022xcs}.
Also, the Julia programming language community
has been developing educational materials that focus on
solving scientific computing problems using parallel and distributed
techniques~\cite{julia2020Covid}.
Most importantly, HPC centers develop web portals to make current HPC applications more 
accessible~\cite{TACCportal},
run short courses to teach others how to use their systems and even how to
use specific applications, and run
hackathons for porting existing applications to new architectures.

\section{Research Questions and Future Directions}

As we move towards implementing an applications-first approach, several
questions warrant investigation:
\begin{enumerate}
\item How might PDC learning outcomes be modified or even rewritten to suit an
applications-first approach?

\item Which applications should be used depending on the student population?
Should proxy applications be used when the actual applications are too large?

\item What pre-requisites are  essential for PDC courses?

\item How can we prepare students to ask effective questions of application domain
experts, HPC system administrators, and even tools like GitHub Co-pilot?

\item What would an "Applications variant" of student
cluster competitions specifically for non-CS students look like? 

\end{enumerate}

\section{Conclusion}

The applications-first approach to teaching parallel and distributed
computing offers a promising path to make these crucial concepts more
accessible and engaging for a broader range of students. By focusing on
practical applications and real-world problem-solving, we can
potentially increase interest in PDC among both CS and non-CS majors.
While challenges remain in implementing this approach, the potential
benefits in terms of increased engagement and broader participation in
PDC education make it a worthy area for further exploration and
development.

\end{document}